\documentstyle[12pt]{article}
\def\fnote#1#2{\begingroup\def\thefootnote{#1}\footnote{#2}
\endgroup}

\begin{document}

\hfill{UTTG-02-97}

\vspace{36pt}

\begin{center}
{\large{\bf Effective Field Theories in the Large $N$
Limit}}

\vspace{36pt}
Steven
Weinberg\fnote{*}{Electronic address:
weinberg@physics.utexas.edu}\\
{\em Theory Group, Department of Physics, University of
Texas\\
Austin, TX, 78712}
\end{center}

\vspace{30pt}

\noindent
{\bf Abstract}
Various effective field theories in four dimensions are
shown to have exact non-trivial solutions in the limit as
the number $N$ of fields of some type becomes large.  These
include extended versions of the $U(N)$ Gross--Neveu model,
the non-linear $O(N)$ $\sigma$-model, and the $CP^{N-1}$
model.  Although these models are not renormalizable in the
usual sense, the infinite number of coupling types allows a
complete cancellation of infinities.  These models provide
qualitative predictions of the form of scattering amplitudes
for arbitrary momenta, but because of the infinite number of
free parameters, it is possible to derive quantitative
predictions only in the limit of small momenta.  For small
momenta the large-$N$ limit provides only a modest
simplification, removing at most a finite number of diagrams
to each order in momenta, except near phase transitions,
where it reduces the infinite number of diagrams that
contribute for low momenta to a finite number.

\vfill

\baselineskip=24pt
\pagebreak
\setcounter{page}{1}

\begin{center}
I. INTRODUCTION
\end{center}

There are a number of instructive models that can be exactly
solved in the limit where the number $N$ of fields becomes
very large.$^1$  Well-known examples include the linear and
non-linear
$\sigma$-models,$^2$ the Gross--Neveu model$^3$  and the
$CP^{N-
1}$ model.$^4$
In four dimensions none of these models except the linear
$\sigma$-model are conventionally renormalizable, so their
large-$N$ limit has usually been
studied either by introducing an ultraviolet cutoff, or by
working in two dimensions, where the simpler versions of
these models are renormalizable.

There is an alternative approach to infinities.   In
effective field
theories that are not renormalizable in the usual
`power-counting' sense, infinities are cancelled
by renormalization of coupling constants provided we
include in the Lagrangian every interaction allowed by
symmetry principles.  Even though this means that the
Lagrangian contains an infinite number of interaction terms,
it is often possible to derive useful results in such
theories by expanding in power of energy rather than
coupling constants.$^5$

In this paper I will show how non-trivial finite results can
be obtained by passing to the limit of large $N$ in various
four-dimensional effective field theories that are not
renormalizable in the conventional sense.  In this task we
will encounter problems both of combinatorics and of
renormalization.

The combinatoric problems here can be illustrated by
recalling the Gross-Neveu model in its original form.
The action is
\begin{equation}
I[\psi]=\int d^2x\;\left[-
\overline{\psi}_r\gamma^\mu\partial_\mu\psi_r-
(g/N)\Big(\overline{\psi}_r\psi_r\Big)^2\right]\;,
\end{equation}
where $\psi_r$ is a set of $N$ fermion fields in two
spacetime dimensions,
forming the defining representation of a $U(N)$ symmetry,
and $g$ is a constant that is held fixed as
$N\rightarrow\infty$.  As an aid to counting factors of
$1/N$,  one cancels the quartic term in Eq.~(1) by adding an
expression that is quadratic in an auxiliary field $\sigma$,
and that vanishes when $\sigma$ is integrated out.   This
results in the replacement of (1) with the equivalent action
\begin{eqnarray}
I[\psi,\,\sigma]&=&I[\psi]+(N/4g)\int
d^2x\left[\sigma+(2g/N)\overline{\psi}_r\psi_r\right]^2
\nonumber\\&=& \int d^2x\;\left[-
\overline{\psi}_r\gamma^\mu\partial_\mu\psi_r+
\sigma\overline{\psi}_r\psi_r+(N/4g)\sigma^2\right]\;.
\end{eqnarray}
Since the fermion field appears quadratically in Eq.~(2) it
may be integrated out, yielding an effective action for
$\sigma$
\begin{equation}
\Gamma[\sigma]=(N/4g)\int d^2x\;\sigma^2-iN\,{\rm
Tr}\,\ln\,\left(\gamma^\mu\partial_\mu-\sigma\right)
\end{equation}
The whole action for $\sigma$ is proportional to
$N$, so
the contribution of graphs with $L$ $\sigma$-loops to
the effective action for $\sigma$ is suppressed by a factor
$N^{1-L}$.  Because this method uses the special properties
of integrals over Gaussians, it is often said that this
method is limited to models in
which the interaction is a product of just two bilinear
currents,$^6$ as in Eq.~(1).

There are also special problems with infinities when an
auxiliary field is introduced in order to impose some
constraint on the $N$-component field, as in the non-linear
$\sigma$-model in four dimensions.  In the original form of
this model the Lagrangian is
\begin{equation}
{\cal L}=-
\mbox{\small$\frac{1}{2}$}f^2\,\partial_\mu\pi_r\,
\partial^\mu\pi_r
\end{equation}
where $f$ is an $N$-independent constant with the dimensions
of mass, and the scalar fields $\pi_r$ form an $O(N)$
$N$-vector,
constrained by
\begin{equation}
\pi_r\pi_r=N\;.
\end{equation}
The counting of powers of $N$ becomes much easier if one
replaces this constraint with a Lagrange multiplier term, so
that the Lagrangian becomes
\begin{equation}
{\cal L}=-
\mbox{\small$\frac{1}{2}$}f^2\,\partial_\mu\pi_r\,
\partial^\mu\pi_r- \mbox{\small$\frac{1}{2}$}
f^2\lambda\left(\pi_r\pi_r-N\right)\;,
\end{equation}
with $\pi_r$ now unconstrained.
Integrating out the auxiliary  field $\lambda(x)$ yields the
Lagrangian (4), with the  $\pi_r$ constrained by Eq.~(5).
If instead we integrate over the $\pi_r$ we find an
effective action for the auxiliary field
\begin{equation}
\Gamma[\lambda]=\frac{iN}{2} {\rm Tr}\,\ln\, (\Box-\lambda)
+\frac{Nf^2}{2}\int d^4x\,\lambda\;.
\end{equation}
Because both terms are proportional to $N$, the Greens
functions for $\lambda(x)$ are given by using the effective
action (7) in the tree approximation.
As well known, this theory is nonrenormalizable.  We can see
this in Eq.~(7).  The field $\lambda(x)$ here has
dimensionality $+2$, so the trace term may be written (aside
from an inconsequential constant term) as
\begin{equation}
\mbox{\small$\frac{1}{2}$}i{\rm Tr}\,\ln (\Box-\lambda)=\int
d^4x\left[{\cal I}_1\lambda+{\cal
I}_2\lambda^2\right]+T_f[\lambda]\;,
\end{equation}
where the ${\cal I}_a$ are divergent constants, and
$T_f[\lambda]$ is finite.  The infinite term ${\cal I}_1$
can be cancelled by an infinity in the parameter $f^2$,
leaving a finite remainder
$\mbox{\small$\frac{1}{2}$}Nf^2_R\int d^4x\;\lambda$ in
$\Gamma(\lambda)$, with
$f_R^2=f^2+2{\cal I}_1$.  But the term ${\cal I}_2$ cannot
be cancelled in this way.  We could of course add a term
proportional to $\lambda^2$ to the Lagrangian  (6), with a
coefficient whose infinite part cancels the infinite
constant ${\cal I}_2$ in Eq.~(8), but then we would lose
the constraint (5), and this would be the linear
$\sigma$-model rather than the nonlinear $\sigma$-model.
If we view this as an effective field theory then the
Lagrangian (4) is just the first term in an infinite series
involving higher powers of the currents $\partial_\mu\phi_r
\partial_\nu\phi_r$ and higher derivatives, but it is not
immediately obvious how these higher terms will allow us to
cancel the infinity in ${\cal I}_2$.

We run into a similar problem when an auxiliary field is
introduced to impose a condition of gauge invariance.  The
classic example here is the $CP^{N-1}$ model in four
dimensions.  This model contains a set of $N$ complex scalar
fields $u_r$, subject to
the constraint that
\begin{equation}
u_r^\dagger(x)\,u_r(x)=N\;.
\end{equation}
In order that the $u_r(x)$ at each $x$ should form a $CP^{N-
1}$ manifold, we must require the action to be invariant
under `gauge' transformations
\begin{equation}
\delta u_r(x)=i\,\epsilon (x)\,u_r(x)\;,
\end{equation}
with $\epsilon(x)$ an arbitrary real infinitesimal function.
In the original $CP^{N-1}$ model, this is accomplished by
taking the action as
\begin{equation}
I=-f^2\int d^4x\;(\partial^\mu u_r-i a^\mu  u_r)^\dagger
(\partial_\mu u_r-i a_\mu  u_r)\;,
\end{equation}
where $f$ is an $N$-independent constant with the dimensions
of mass, and $a_\mu(x)$ may be defined as the bilinear
\begin{equation}
a_\mu \equiv  -(i/2N)\Big(u_r^\dagger \partial_\mu u_r-
(\partial_\mu u_r^\dagger) u_r\Big)
\end{equation}
which under the gauge transformation (10) changes by
\begin{equation}
\delta a_\mu=\partial_\mu\epsilon\;.
\end{equation}
Equivalently, we can replace $a_\mu(x)$ in Eq.~(11) with an
independent
auxiliary field $A_\mu(x)$, so that the action is
\begin{equation}
I=-f^2\int d^4x\;(\partial^\mu u_r-i A^\mu  u_r)^\dagger
(\partial_\mu u_r-i A_\mu  u_r)\;.
\end{equation}
Since $A_\mu(x)$ enters quadratically in Eq.~(14),
the path integral over $A_\mu(x)$ is done by giving it a
value at which the action (14) is stationary with respect to
$A_\mu(x)$, which turns out to give $A_\mu(x)=a_\mu(x)$.  To
enforce
the constraint (9) we can add a Lagrange multiplier term
$-f^2\int d^4x\,\lambda\,(u_r^\dagger u_r-N)$, with
$\lambda(x)$ another auxiliary field, and $u_r(x)$ now
unconstrained.  Since $u_r(x)$ enters quadratically in the
action it can be integrated out, yielding an effective
action for the auxiliary fields
\begin{equation}
\Gamma[A,\lambda]=iN\,{\rm Tr}\,\ln \Big[D_\mu D^\mu-
\lambda\Big]+Nf^2\int d^4x\;\lambda\;,
\end{equation}
where here $D_\mu\equiv\partial_\mu-i A_\mu$.  Because each
term is proportional to $N$, the contribution of graphs with
$L$ loops to the Greens functions for $\lambda$ and $a_\mu$
is suppressed by a factor $N^{1-L}$.  Eq.~(15) displays the
problem with renormalizabilty in four dimensions:
Dimensional analysis and gauge invariance show that the
infinite part of the trace is a linear combination of $\int
d^4x\, \lambda$, $\int d^4x\, \lambda^2$, and $\int
d^4x\,(\partial_\mu A_\nu-\partial_\nu A_\mu)^2$ with
divergent coefficients, and although the infinite part of
the coefficient of $\int d^4x\, \lambda$ can be cancelled by
an infinite term in $f^2$, there is nothing here that can
cancel the infinite coefficients of $\int d^4x\, \lambda^2$
or $\int d^4x\,(\partial_\mu A_\nu-\partial_\nu A_\mu)^2$.
Treating this as an effective field theory, we would
certainly have to add terms to the action involving
$\int d^4x\,(\partial_\mu a_\nu-\partial_\nu a_\mu)^2$,
where $a_\mu$ is defined by Eq.~(12), but this would not
cancel the infinity in $\Gamma[\lambda,A]$ proportional to
$\int d^4x\,(\partial_\mu A_\nu-\partial_\nu A_\mu)^2$.  We
cannot add a term proportional to $\int d^4x\,(\partial_\mu
A_\nu-\partial_\nu A_\mu)^2$ without making $A_\mu(x)$ an
independent dynamical field, thus removing the most
interesting aspect of the theory, the appearance of
long-range forces in a theory without an elementary gauge
field.

Finally, there is a problem that always confronts us in
dealing with effective field theories: how to use a theory
with an infinite number of free parameters to
derive physical predictions.  As we shall see, the large $N$
limit can give qualitative information about the {\em form}
of $S$-matrix elements, but in effective field theories it
is not possible actually to calculate the functions that
appear in $S$-matrix elements except in the low energy
limit.  In the extended Gross--Neveu model considered in
Section II it turns out that there are usually only a finite
number of graphs that contribute to each order in energy,
whatever the value of $N$, so for low energy the large $N$
limit leads only to modest simplifications.   As shown in
Sections III and IV, the same is true in the extended
non-linear $\sigma$-model and the extended $CP^{N-1}$ model,
with one interesting exception:  Near the phase transitions
at which the broken symmetries of these models are restored,
there is an infinite number of graphs of the same order in
energy, which can be summed only in the large $N$ limit.

The original motivation of this work was to decide whether
 the appearance of a spin-one `photon' in the
two-dimensional $CP^{N-1}$ model occurs also in
four-dimensional versions of this model, when the problem
of infinities is handled by treating the model as
an effective quantum field theory.  Section IV shows
that the answer to this question is yes, but as discussed
in Section V, this result is less surprising than
might be supposed.

\vspace{12pt}
\begin{center}
II. THE EXTENDED GROSS--NEVEU MODEL
\end{center}

\nopagebreak
To illustrate the use of the large $N$ limit in effective
field theories, let us consider the general class of models
with
a set of $N$ massless fermion fields in $d$ spacetime
dimensions, transforming according to the defining
representation of a global $U(N)$ symmetry.   Any
$U(N)$-invariant action will be a functional of a set of
bilinear currents $j_\ell(x)$ that are invariant under
$U(N)$, such as the currents $j_0=\overline{\psi}_r\psi_r$,
$j_{1\mu}=\overline{\psi}_r\partial_\mu\psi_r$, etc.  We
will consider a class of extended four-dimensional
Gross--Neveu models, with an action of the
form\fnote{**}{The free-field action could itself be
regarded as a linear term in $NF[j/N]$, but it is convenient
to treat it separately.}
\begin{equation}
I[\psi]=-\int
d^dx\;\overline{\psi}_r\gamma^\mu\partial_\mu\psi_r+NF[j/N]
\;,
\end{equation}
where  $F[{\tau}]$  is some $N$-independent functional.  The
original Gross--Neveu action (1) is a special case of
Eq.~(16), with $F[j/N]$ quadratic in the particular current
$j_0=\overline{\psi}_r\psi_r$.  As we shall see, the
$N$-dependence given the second term in the action
(16) makes the theory non-trivial but soluble, as in the
original Gross--Neveu model.

The action (16) may be replaced with an equivalent action
\begin{equation}
I[\psi,\sigma]=\int d^4x\;\left[-
\overline{\psi}_r\gamma^\mu\partial_\mu\psi_r+
\sigma_\ell(x)j_\ell(x)\right]+NG[\sigma,N]
\end{equation}
where $\exp\{iNG[\sigma,N]\}$ is the functional Fourier
transform with respect to $N\tau$ of $\exp\{iNF[\tau]\}$
\begin{equation}
\exp\left\{iNG[\sigma,N]\right\}\equiv
\int\prod_{\ell,x}d\tau_\ell(x)\;
\exp\left\{-iN\int
d^4x\,\sigma_\ell(x)\tau_\ell(x)+iNF[\tau]\right\}\;.
\end{equation}
Of course, $\exp\{iNF[\tau]\}$ is then (up to an unimportant
constant factor) the Fourier transform of
$\exp\{iNG[\sigma,N]\}$
\begin{equation}
\exp\left\{iNF[\tau]\right\}\propto
\int\prod_{\ell,x}d\sigma_\ell(x)\;
\exp\left\{iN\int
d^4x\,\sigma_\ell(x)\tau_\ell(x)+iNG[\sigma,N]\right\}\;.
\end{equation}
The integral over the $\sigma_\ell(x)$ in the functional
integral of $\exp\{iI[\psi,\sigma]\}$ thus yields
\begin{equation}
\int \prod_{\ell,x}d\sigma_\ell(x)\;\exp\{iI[\psi,\sigma]\}
\propto\exp\{iI[\psi]\}
\end{equation}
so the action $I[\psi,\sigma]$ given by Eq.~(17) is
equivalent to the original action (16).

In the limit of large $N$ the Fourier integral (18) may be
done by setting $\tau_\ell(x)$ at the stationary `point'
$\tau^\sigma(x)$ of the integrand, at which
\begin{equation}
\left.\frac{\delta
F[\tau]}{\delta\tau_\ell(x)}\right|_{\tau=\tau^\sigma}\equiv
\sigma_\ell(x)\;.
\end{equation}
so that $G[\sigma,N]$ approaches an $N$-independent
functional, the Legendre transform of $F[\tau]$:
\begin{equation}
G[\sigma,N]\rightarrow G[\sigma]\equiv F[\tau^\sigma]-\int
d^dx\;\sigma_\ell(x)\tau^\sigma_\ell(x)\;,
\end{equation}
That is, for $N\rightarrow\infty$, the action may be taken
as
\begin{equation}
I[\psi,\sigma]=\int d^4x\;\left[-
\overline{\psi}_r\gamma^\mu\partial_\mu\psi_r+
\sigma_\ell(x)j_\ell(x)\right]+NG[\sigma]\;.
\end{equation}
We could just as well have taken the action (23) as our
starting point, with $G[\sigma]$ an arbitrary
$N$-independent functional; the only difference would be
that
then the theory would be equivalent to one with an action of
the original form (16), but with $F[\tau]$ independent of
$N$ only in the limit $N\rightarrow\infty$.

Now, we want to calculate the quantum effective action
$\Gamma[\psi,\sigma]$, which by definition in the tree
approximation gives
the same result as the sum of all loop and tree graphs
calculated using the action (23).  According to the usual
prescription, we must replace $\psi_r(x)$ and
$\sigma_\ell(x)$ in Eq.~(23) with sums
$\psi_r(x)+\psi'_r(x)$
and $\sigma_\ell(x)+\sigma'_\ell(x)$ and integrate over the
quantum perturbations $\psi'_r(x)$ and $\sigma'_\ell(x)$,
including only one-particle-irreducible graphs.  A standard
power-counting argument gives, to leading order in $1/N$,
\begin{equation}
\Gamma[\psi, \sigma]\rightarrow\int d^4x\;\left[-
\overline{\psi}_r\gamma^\mu\partial_\mu\psi_r+
\sigma_\ell(x)j_\ell(x)\right]+NT[\sigma]+NG[\sigma]\;,
\end{equation}
where $T[\sigma]$ is an $N$-independent functional of
$\sigma$, like the functional ${\rm
Tr}\,\ln(\gamma^\mu\partial_\mu-\sigma)$ in Eq.~(3),
defined in general by the integral of a Gaussian
\begin{equation}
\exp\left\{iNT[\sigma]\right\}\equiv\int\left[\prod_{n,x}
d\psi'_r(x)\right]
\exp\left\{i\int d^dx\;\left[-
\overline{\psi}'_r\gamma^\mu\partial_\mu\psi'_r+
\sigma_\ell(x)\,j'_\ell(x)\right]\right\}\;.
\end{equation}
[To obtain Eq.~(24), note first that purely fermionic loops
yield
a term $NT[\sigma]$ in the effective action, which just
makes an
additive contribution to $NG[\sigma]$.  The $\sigma_\ell$
propagators are then given by the inverse of the coefficient
of
the quadratic term in $NT[\sigma]+NG[\sigma]$, and are hence
proportional to $1/N$, while the purely bosonic vertices
(including those derived from $T[\sigma]$) make
contributions proportional to $N$.  Thus a graph with
$V_\sigma$ purely bosonic vertices, $V_\psi$
fermion--fermion--boson vertices, $I_\psi$ internal fermion
lines (excluding those in purely fermionic loops) and
$I_\sigma$ internal boson lines makes a contribution of
order $N^{V_\sigma-I_\sigma}$.  But $2V_\psi=2I_\psi+F$,
where $F$ is the number of external fermion lines (i.e.,
factors of the classical $\psi$ field), and the number $L$
of loops (with all purely fermionic loops counted as bosonic
vertices) is $L=I_\psi+I_\sigma-V_\psi-V_\sigma+1$, so
the number of powers of $N$ is $V_\sigma-I_\sigma=1-L-F/2$.
The leading graphs are
therefore those with no loops, but the only
one-particle-irreducible graphs with no loops are those
consisting of just a single vertex, which yield the result
(24).]

For instance, the amplitude for fermion--fermion scattering
is given to leading order in $1/N$ by the tree graphs in
which a single $\sigma$ line is exchanged between the
external fermions, with the vertices given by the term $\int
d^4x\,\sigma_\ell(x)j_\ell(x)$ in Eq.~(24), and the $\sigma$
propagator given by the inverse of the coefficient of the
quadratic term in $NG[\sigma]+NT[\sigma]$.
Unlike the case of the original Gross--Neveu model, the
terms of higher order in $1/N$ arise not only from loop
graphs that correct the effective action (24), but also
from higher-order terms in $G[\sigma,N]$.

Now let's take up the problem of renormalization in the
large $N$ limit.  Although the one-loop functional
$T[\sigma]$ is
highly non-local, its infinite part is `perturbatively
local' --- that is, it is the integral of a series (in
general infinite) of products of fields and their
derivatives with divergent coefficients, of which only a
finite number of terms contribute to the tree amplitude for
any given process to any finite order in momentum.  In order
to cancel these infinities, it is necessary that $G[\sigma]$
be a general perturbatively local functional, subject to no
constraints other than symmetry properties that also
constrain $T[\sigma]$.  But any perturbatively local
$G[\sigma]$  can be obtained as the Legendre transform
of some
perturbatively local $F[\tau]$, provided only that
\begin{equation}
{\rm Det}{\cal M}\neq 0\;,~~~~~~~~~~{\cal
M}_{\ell\,x\,,m\,y}\equiv\frac{\delta^2 G[\sigma]}{\delta
\sigma_\ell(x)\delta \sigma_m(y)}\;.
\end{equation}
  [To see this, it is only
necessary to note that the functional $G[\sigma]$ is the
Legendre transform of a
functional $F[\tau]$ given by the inverse Legendre
transform of $G[\sigma]$:
\begin{equation}
F[\tau]=G[\sigma^\tau]+\int
d^dx\;\sigma_\ell^\tau(x)\,\tau_\ell(x)
\end{equation}
with $\sigma^\tau$ the stationary `point' of the expression
on the right-hand side
\begin{equation}
\left.\frac{\delta
G[\sigma]}{\delta\sigma_\ell(x)}\right|_{\sigma=\sigma^\tau}
=-\tau_\ell(x)\;.
\end{equation}
In terms of Feynman graphs, this just says that $F[\tau]$ is
given a sum of tree Feynman graphs calculated from the
action $G[\sigma]+\int d^dx\,\sigma_\ell(x)\tau_\ell(x)$,
for which the propagator is ${\cal M}^{-
1}_{\ell\,x\;,m\,y}$.  Therefore $F[\tau]$  is
perturbatively local if $G[\sigma]$ is, and if ${\rm
Det}{\cal M}\neq 0$.]
Furthermore, the condition that  ${\rm Det}{\cal M}\neq 0$
can always be satisfied by adding a finite quadratic term
to $G[\sigma]$, and subtracting the same term from
$T[\sigma]$.  Apart from symmetries, the only
constraint on $F[\tau]$ is that it be perturbatively local,
so it is always possible to choose an $F[\tau]$ that gives
whatever $G[\sigma]$ is needed to cancel the infinities in
$T[\sigma]$.

This works out in a particularly simple way if  the currents
$j_\ell(x)$ have dimensionality (in powers of mass, with
$\hbar=c=1$) less than the spacetime dimensionality, so
that  the $\sigma_\ell(x)$ have positive dimensionality.  In
this case the infinite part $T_\infty[\sigma]$ of
$T[\sigma]$ is the integral of a {\it polynomial} in the
$\sigma_\ell(x)$ and their derivatives, with infinite
constant coefficients.  For instance, consider an extended
Gross--Neveu model in four dimensions, with an action of the
form
\begin{equation}
I[\psi]=-\int
d^4x\;\overline{\psi}_r\gamma^\mu\partial_\mu\psi_r+NF[j_0/N
]\;,
\end{equation}
where  $F[j_0/N]$ is an arbitrary even local functional of
the
single current
\begin{equation}
j_0=\overline{\psi}_r\psi_r\;.
\end{equation}
We take $F[j/N]$ even so that the action will be invariant
under a discrete chiral symmetry transformation
$\psi_r\rightarrow\gamma_5\psi_r$, which if unbroken keeps
the fermions massless.
Here $\sigma(x)$ has dimensionality $+1$, and chiral
symmetry tells us that the functional $T[\sigma]$ defined by
Eq.~(25) is even in $\sigma$, so this functional takes the
form
\begin{equation}
T[\sigma]=\int d^4x\;\left[{\cal I}_0+{\cal
I}_1\sigma^2+{\cal I}_2\sigma\Box\sigma
+{\cal I}_3\sigma^4\right]+T_f[\sigma]
\end{equation}
where the ${\cal I}_a$ are infinite constants, and $T_f$ is
a finite functional, which for constant $\sigma$ takes the
well-known form$^7$
$$
T_f[\sigma]=-\frac{1}{32\pi^2}\int d^4x\;
\sigma^4\ln\sigma^2\;.
$$
  The functional $F[\tau]$ may be expanded in a series of
even local operators of increasing dimensionality
\begin{equation}
F[\tau]=\int d^4x\;\left[A_0+A_1\tau^2+A_2\tau \Box\tau
+A_3\tau^4+A_4\tau\Box\Box\tau+\cdots\right]\;,
\end{equation}
in which case Eqs.~(21) and (22) give
\begin{equation}
G[\sigma]=\int d^4x\;\Bigg[A_0 -
\frac{1}{4A_1}\sigma^2+\frac{A_2}{4A_1^2}\sigma\Box\sigma
+\frac{A_3}{16A_1^4}\sigma^4+\left(\frac{A_4}{4A_1^2}-
\frac{A_2^2}{4A_1^3}\right)\sigma\Box\Box\sigma+\cdots\Bigg]
\;.
\end{equation}
In order to cancel infinities, we must take
the bare parameters as
\begin{eqnarray}
&&A_0=C_0- {\cal I}_0\;,~~~~ A_1=-
\mbox{\small$\frac{1}{4}$}[C_1- {\cal I}_1]^{-
1}\;,~~~~A_2=\frac{C_2- {\cal I}_2}{4[{\cal C}_1-
{\cal I}_1]^{2}}\;,
\nonumber\\&&
A_3=\frac{C_3- {\cal I}_3}{16[C_1- {\cal I}_1]^4}\;,~~~~
A_4=\frac{C_4}{4[C_1- {\cal I}_1]^{2}}-\frac{[C_2- {\cal
I}_2]^2}{4[C_1- {\cal I}_1]^{3}}\;,~~~\cdots
\nonumber\\&&
\end{eqnarray}
where the $C_a$ are the finite renormalized coupling
parameters that appear in the final result
\begin{equation}
G[\sigma]+T[\sigma]=\int d^4x\;\Bigg[C_0 +
C_1\sigma^2+C_2\sigma\Box\sigma
+C_3\sigma^4+C_4\sigma\Box\Box\sigma+\cdots\Bigg]+T_f[\sigma
]\;.
\end{equation}
It is important but not surprising that the infinite number
of unrenormalized constants $A_a$ can be chosen to give
finite results for $\Gamma[\sigma]$, despite the fact that
the Gross--Neveu model is not conventionally renormalizable
in four dimensions.  What is somewhat surprising is that
this is possible with an interaction given by a power series
in the single current $j_0=\overline{\psi}_r\psi_r$ and its
derivatives, without needing to include additional currents,
such as $j_{1\mu}\equiv\overline{\psi}_r\partial_\mu\psi_r$.
Although it is not necessary to include additional currents
in order to cancel infinities, in the spirit of effective
field theory we really should include all $U(N)$-invariant
currents in the action.  This complicates the cancellation
of infinities through renormalization, but as we have shown
earlier,
it does not make it impossible.

   What good is a theory like this, that has an infinite
number of arbitrary parameters?  For an action of the form
(29), the effective action (24) takes the form
\begin{equation}
\Gamma[\psi,\sigma]=-\int
d^4x\;\overline{\psi}_r\gamma^\mu\partial_\mu\psi_r
+\int d^4x\; \sigma \overline{\psi}_r\psi_r
+N
G[\sigma]+NT[\sigma]
\;,
\end{equation}
with $G[\sigma]+T[\sigma]$ given by Eq.~(35).  To calculate
scattering amplitudes we must use this as the action in the
tree approximation.  For instance, the invariant amplitude
for a fermion--fermion scattering process $A+B\rightarrow
C+D$ takes the form
\begin{eqnarray}
M(A+B\rightarrow
C+D)&=&\delta_{r_Ar_C}\delta_{r_Br_D}\,(\bar{u}_Cu_A)\,
(\bar{u}_Du_B)\, \Delta(t)\nonumber\\&&-
\delta_{r_Ar_D}\delta_{r_Br_C}\,(\bar{u}_Du_A)\,(\bar{u}_Cu_
D)\,\Delta(u)\;,
\end{eqnarray}
where $t$ and $u$ are the Mandelstam variables  $t=-
(p_A-p_C)^2$ and $u=-(p_A-p_D)^2$, and
$\Delta(t)$ is the $\sigma$ propagator.  This particular
form of the scattering amplitude is a consequence of the
assumption that the action has the form (29), with only the
one current $j_0$, and it is
valid in the large $N$ limit for
arbitrary values of the fermion momenta.

Unfortunately, to go further and actually calculate the
propagator $\Delta(s)$ without knowing the infinite number
of free parameters in $F[\tau]$ or $G[\sigma]$, it is
necessary to take the limit of small momenta.  But in this
limit, little is gained by also letting $N$ become large.

To compare the consequences of the large $N$ and low
momentum approximations, let us consider an amplitude
calculated using the action (29) in the equivalent form
\begin{equation}
I[\psi,\sigma]=-\int
d^4x\;\overline{\psi}_r\gamma^\mu\partial_\mu\psi_r
+\int d^4x\; \sigma \overline{\psi}_r\psi_r
+NG[\sigma]
\;,
\end{equation}
with the $N$-dependence of $G[\sigma]$ left unspecified.
Take all incoming and outgoing momenta of the order of some
small momentum $Q$, and suppose that infinities are
cancelled by renormalizing at momenta also of order $Q$.
Fermion propagators go as $Q^{-1}$, while $\sigma$
propagators go as constants (any quadratic terms in
$G[\sigma]$ involving derivatives being treated as
interactions) and each loop introduces four factors of $Q$,
so an amplitude with $I_\psi$ internal fermion lines and $L$
loops goes as $Q^\nu$, where
\begin{equation}
\nu=4L-I_\psi+\sum_id_iV_i\;,
\end{equation}
where $V_i$ is the number of purely bosonic interactions of
type $i$, and $d_i$ is the number of derivatives in such an
interaction.  This may be rewritten by using the familiar
formulas
\begin{equation}
2I_\psi+E_\psi=2V_\psi\;,
\end{equation}
\begin{equation}
2I_\sigma+E_\sigma=\sum_iV_in_i+V_\psi\;,
\end{equation}
and
\begin{equation}
L=I_\psi+I_\sigma-\sum_iV_i-V_\psi+1\;,
\end{equation}
where $I_\sigma$ is the number of internal $\sigma$ lines,
$E_\psi$ and $E_\sigma$ are the numbers of external fermion
and $\sigma$ lines, $V_\psi$ is the number of
fermion--fermion--$\sigma$ vertices, and $n_i$ is the number
of
$\sigma$ fields in the purely bosonic interaction of type
$i$.
Eq.~(39) then  may be put in the form
\begin{equation}
\nu=2L+\sum_i\Delta_iV_i+2-\frac{E_\psi}{2}-
E_\sigma+2\;,
\end{equation}
where
\begin{equation}
\Delta_i=n_i+d_i-2\;.
\end{equation}
Now, all $\sigma$ interactions have either $n_i\geq 4$ or
$n_i=2$ and $d_i\geq 2$ (the term proportional to $\int
\sigma^2\,d^4x$ in $G[\sigma]$ being treated as the
kinematic action for the $\sigma$ field), so for all purely
bosonic interactions $\Delta_i\geq 2$.
A scattering amplitude for a fixed process (i.e., $E_\psi$
and $E_\sigma$ fixed) is therefore dominated for
$Q\rightarrow 0$ by {\em tree} graphs with any number of
fermion--fermion--$\sigma$ vertices but no loops and no pure
$\sigma$ interactions.   These graphs are a subset of those
we encountered in the limit of large $N$, so the large $N$
limit would introduce no further simplification here.

The large $N$ limit becomes relevant in the next order in
$Q^2$, which according to Eqs.~(43) and (44) is given by
graphs with any number of fermion--fermion--$\sigma$
vertices and either one loop or with no loops and one pure
$\sigma$ interaction with $\Delta_i=2$.  The graphs with a
single interaction from $G[\sigma]$ or a single purely
fermionic loop reproduce what we would find by using the
large $N$ effective action (36) to this order in $Q^2$.  But
here there is also a class of graphs that do not appear in
the large $N$ limit: the graphs constructed out of only
fermion--fermion--$\sigma$ vertices and containing a single
loop, which is not purely fermionic.  (For instance, in
fermion--fermion scattering these are the graphs in which a
pair of $\sigma$
lines are exchanged between the fermions.)  The   large $N$
limit is therefore useful here, but not very useful, because
even without it there are only a finite number of graphs to
each order in $Q^2$.  This is just a consequence of the fact
that this is a theory where all interactions are
nonrenormalizable.  In the next section we will see an
example where the large $N$ limit is much more important,
because there is an infinite number of graphs to each order
in small momenta, which can not be summed except in the
limit of large $N$.

\vspace{12pt}
\begin{center}
III. THE NON-LINEAR $\sigma$-MODEL: INTEGRATING IN AN ORDER
PARAMETER
\end{center}
\nopagebreak

Auxiliary fields are sometimes introduced to enforce
constraints on the other fields, as well as to help in
counting factors of $1/N$.  The classic example is the
non-linear $O(N)$ $\sigma$-model, which in its usual form
has Lagrangian (4).  Here we will consider a class of
extended non-linear $\sigma$-models, with Lagrangians of the
form
\begin{equation}
I[\pi]=-\frac{f^2}{2}\int
d^4x\;\partial_\mu\pi_r\partial^\mu\pi_r+Nf^2F[j/N]\;,
\end{equation}
where  $\pi_r$ is a set of $N$ scalar fields, satisfying the
constraint
\begin{equation}
\pi_r\pi_r=N\;;
\end{equation}
$j(x)$ is the $O(N)$-invariant scalar current with the
minimum number of derivatives
\begin{equation}
j\equiv \mbox{\small$\frac{1}{2}$}\partial_\mu\pi_r
\partial^\mu\pi_r\;;
\end{equation}
$f^2$ is an arbitrary positive constant; and
$F[\tau]$ is a functional that, apart from being
perturbatively local and $N$-independent, can be chosen as
we like.  The $N$-dependence in Eq.~(45) has been chosen so
that this model will be soluble but non-trivial in the limit
$N\rightarrow\infty$.

As in the case of the extended Gross--Neveu model, we shall
introduce an auxiliary field $\sigma(x)$, and replace
Eq.~(45) with the equivalent action
\begin{equation}
I[\pi,\sigma]=Nf^2G[\sigma,N]-f^2\int d^4x\;(1+\sigma)\,
j\;,
\end{equation}
where $\exp\{iNf^2G[\sigma,N]\}$ is the functional Fourier
transform with respect to $f^2N\tau$ of
$\exp\{iNf^2F[\tau]\}$:
\begin{equation}
\exp\left\{iNf^2G[\sigma,N]\right\}\equiv
\int\prod_{x}d\tau(x)\;
\exp\left\{iNf^2\int
d^4x\,\sigma(x)\tau(x)+iNf^2F[\tau]\right\}\;.
\end{equation}
It is easy to see that we get back to Eq.~(45) when we
integrate out $\sigma(x)$, but it will be convenient instead
to use the action in the form (48).

In the limit of large $N$ the functional $G[\sigma,N]$
approaches an $N$-independent functional given by the
Legendre transform of $F[\tau]$
\begin{equation}
G[\sigma,\tau]\rightarrow G[\sigma]=\int
d^4x\;\sigma\tau^\sigma+F[\tau^\sigma]
\end{equation}
with $\tau^\sigma$ defined by
\begin{equation}
\left.\frac{\delta
F[\tau]}{\delta\tau(x)}\right|_{\tau=\tau^\sigma}=-
\sigma(x)\;.
\end{equation}

At this point, it is not so obvious how infinite
counterterms in the functional $G[\sigma]$ can be used to
cancel the ultraviolet divergence proportional to $\int
d^4x\,\lambda^2$  that is encountered when we introduce a
Lagrange
multiplier $\lambda(x)$ and integrate over the $\pi_r$.  Yet
we know that this is possible, because the cancellation of
infinities is obvious in an extended linear $\sigma$ model,
in which the action is an arbitrary perturbatively local
functional of an unconstrained $N$-vector field $\phi_r$,
and from such a model it is always possible to construct a
non-linear $\sigma$-model by integrating out the massive
order-parameter represented by the $O(N)$ scalar
$\sqrt{\phi_r\phi_r}$.  This suggests that we should show
the cancellation of infinities in the extended non-linear
$\sigma$ model by using the ingredients appearing in
Eq.~(48) to construct something like a linear
$\sigma$-model.

For this purpose, let us define new fields
\begin{equation}
\phi_r\equiv f\sqrt{1+\sigma}\,\pi_r\;.
\end{equation}
Using the constraint (46), the current (47) may be written
\begin{equation}
j=\frac{1}{2f^2(1+\sigma)}\partial_\mu\phi_r
\partial^\mu\phi_r- \frac{N}{8(1+\sigma)^2}\,\partial_\mu
\sigma\;\partial^\mu \sigma\;.
\end{equation}
Also, the constraint now reads
\begin{equation}
\phi_r\phi_r=Nf^2(1+\sigma)
\end{equation}
and will again be imposed by introducing a Lagrange
multiplier $\lambda(x)$.  The action (48) is thereby
replaced with the equivalent action
\begin{equation}
I[\phi,\sigma,\lambda]
=Nf^2G'[\sigma]-\int
d^4x\;\Bigg[\mbox{\small$\frac{1}{2}$}\lambda\Big(\phi_r\phi
_r-
Nf^2(1+\sigma)\Big)+\mbox{\small$\frac{1}{2}$}\partial_\mu
\phi_r
\partial^\mu\phi_r\Bigg]\;.
\end{equation}
where
\begin{equation}
G'[\sigma]\equiv G[\sigma]+\frac{1}{8}\int d^4x
\frac{\partial_\mu\sigma\partial^\mu\sigma}{1+\sigma}
\end{equation}
Now it is $\lambda(x)$ rather than $\sigma(x)$ that
interacts with the $N$-vector of scalar fields, so the
$\phi_r$ scattering amplitude may be calculated in terms of
the effective action  for $\lambda(x)$ and $\phi_r(x)$,
which is obtained by integrating out $\sigma(x)$.  The part
of the action involving $\sigma(x)$ is proportional to $N$,
and does not involve the $\phi_r(x)$, so we can integrate
out $\sigma(x)$ to leading order in $1/N$ by setting
$\sigma(x)$ equal to the value $\sigma^\lambda(x)$ where
(55) is stationary with respect to $\sigma(x)$:
\begin{equation}
\left.\frac{\delta G'[\sigma]}{\delta \sigma(x)}
\right|_{\sigma=\sigma^\lambda}=-
\mbox{\small$\frac{1}{2}$}\lambda(x)\;.
\end{equation}
This gives an  action for $\phi_r(x)$ and $\lambda(x)$:
\begin{equation}
I[\phi,\lambda]=NH[\lambda]-\int
d^4x\;\left[\mbox{\small$\frac{1}{2}$}\partial_\mu\phi_r
\partial^\mu\phi_r+\mbox{\small$\frac{1}{2}$}\lambda\phi_r
\phi_r\right]\;,
\end{equation}
where $H[\lambda]$ is another Legendre transform
\begin{equation}
H[\lambda]\equiv f^2G'[\sigma^\lambda]+
\mbox{\small$\frac{1}{2}$}f^2\int
d^4x\;\lambda\,(1+\sigma^\lambda)\;.
\end{equation}
The same reasoning as in the previous section (with
$\lambda$ and $\pi_r$ replacing $\sigma$ and $\psi_r$) shows
that to leading order in $1/N$, the quantum effective action
which in tree approximation gives the complete scattering
amplitude is here
\begin{equation}
\Gamma[\phi,\lambda]=\Gamma[\lambda]-\int
d^4x\;\left[\mbox{\small$\frac{1}{2}$}\partial_\mu\phi_r
\partial^\mu\phi_r+\mbox{\small$\frac{1}{2}$}\lambda\phi_r
\phi_r\right]\;.
\end{equation}
where
 \begin{equation}
\Gamma[\lambda]=NH[\lambda]+\mbox{\small$\frac{1}{2}$}iN
{\rm Tr}\,\ln (\Box-\lambda)\;.
\end{equation}
Following the same argument as in the previous section, by
choosing $F[\tau]$ we can make $G'[\sigma]$ and $H[\lambda]$
any perturbatively local functionals we like, so we can
adjust $H[\lambda]$ to cancel the infinite terms in the
one-loop trace ${\rm Tr}\,\ln (\Box-\lambda)$ proportional
to
$\int d^4x\,\lambda$ and $\int d^4x\,\lambda^2$.  For
$\lambda$ spacetime-independent, this gives
\begin{equation}
V(\lambda)=\frac{N\lambda^2\,\ln(\lambda/M^2)}{64\pi^2}+N
\sum_{n=0}^\infty c_n\lambda^n
\end{equation}
where the $c_n$ are model-dependent $N$-independent finite
constants; $M$ is
a constant that can be chosen for instance so that $c_2=1$;
and $V[\lambda]$ is the `effective potential,' defined so
that for a spacetime-independent $\lambda$,
\begin{equation}
\Gamma[\lambda]=-{\cal V}_4\,V(\lambda)\;,
\end{equation}
with ${\cal V}_4$ the spacetime volume.

To identify the possible phases of this model, we must
examine the possible spacetime-independent vacuum
expectation values  of the fields $\phi_r$ and $\lambda$,
at which the effective action (60) is stationary.  These
phases are of two different types:

\vspace{9pt}
\begin{center}
{\em 1. Broken Symmetry Phase}
\end{center}
In this phase the vacuum expectation value of $\lambda(x)$
vanishes, while $\phi_r(x)$ has a vacuum expectation
$\sqrt{N}v_r$, given by the solution of the equation
\begin{equation}
\left.\frac{\delta\Gamma[\lambda]}{\delta
\lambda(x)}\right|_{\lambda(x)=0}=-\left.\frac{\partial
V(\lambda)}{\partial\lambda}\right|_{\lambda=0}=\mbox{\small
$\frac{1}{2}$}Nv_rv_r\;.
\end{equation}
From Eqs.~(62) and (64) we see the system will be in this
phase  if $c_1<0$, and that in this case
$v_r$ is $N$-independent and given by
\begin{equation}
v_rv_r=-2c_1\;.
\end{equation}
To see the particle content of the theory in this phase,
we note that the terms in the effective action (60) of
second order in the displacement of the fields from their
equilibrium value have a coefficient matrix given in
momentum space by $-\mbox{\small$\frac{1}{2}$}{\cal D}(k)$,
where ${\cal D}(k)$ is the matrix
\begin{eqnarray}
{\cal D}_{rs}(k)&=&k^2\,\delta_{rs}\;,~~~~~~~~ {\cal
D}_{\lambda\lambda}(k)~=~NA(k^2)\;, \nonumber\\
{\cal D}_{r\lambda}(k)&=&{\cal D}_{\lambda
r}(k)~=~\sqrt{N}v_r\;.
\end{eqnarray}
Here $A(k^2)$ is an $N$-independent one-loop amplitude
derived from the part of $\Gamma[\lambda]$ quadratic in
$\lambda$, which is easily calculated to be
\begin{equation}
A(k^2)=\frac{\ln (k^2/{\cal
M}^2)}{32\pi^2}+\sum_{n=1}^\infty d_n (k^2)^n
\end{equation}
with $d_n$ another set of model-dependent finite constants,
and ${\cal M}$ is a constant chosen so that the constant
term $d_0$ in the sum is absent.
The scalar propagator $\Delta(k)={\cal D}^{-1}(k)$ hence has
elements
\begin{eqnarray}
\Delta_{rs}(k)=\frac{1}{k^2}\left(\delta_{rs}-
\frac{v_rv_s}{v^2-k^2A(k^2)}\right)\;&,&~~~
\Delta_{\lambda\lambda}(k)~=~-\frac{k^2}{N(v^2-
k^2A(k^2))}\;, \nonumber\\
\Delta_{r\lambda}(k)=\Delta_{\lambda
r}(k)&=&\frac{v_r}{\sqrt{N}(v^2-k^2A(k^2))}\;,
\end{eqnarray}
where $v^2\equiv v_rv_r$.  The pole in $\Delta_{rs}$ at
$k^2=0$ clearly arises from the Goldstone bosons of
$O(N)/O(N-1)$.  This pole occurs only in the propagators of
the components of $\phi_r$ in directions perpendicular to
$v_r$.  There may be other particles associated with poles
at model-dependent non-zero masses arising from the
vanishing of the denominators $v^2-k^2A(k^2)$ in the
propagators of the fields $v_r\phi_r$ and $\lambda$, but
without special assumptions about $H[\lambda]$ we can say
nothing about them, except that they do not mix with the
Goldstone bosons.  The invariant amplitude for
Goldstone boson--Goldstone boson scattering is given by the
$\lambda$--$\lambda$ element of the propagator, as
\begin{equation}
M(ap,bq\rightarrow
a'p',b'q')=\frac{1}{N}\Bigg[\frac{\delta_{ab}\delta_{a'b'}s}
{v^2+s A(-s)}+\frac{\delta_{aa'}\delta_{bb'}t}{v^2+t A(-
t)}+\frac{\delta_{ab'}\delta_{a'b}u}{v^2+u A(-u)}\Bigg]\;,
\end{equation}
where $a$, $b$, $a'$, $b'$ run over the Goldstone
directions, from 1 to $N-1$ (with $v_r$ taken in the
$N$-direction), and $s$, $t$, $u$ are the usual Mandelstam
variables: $s=-(p+q)^2$, $t=-(p-p')^2$, and $u=-(p-q')^2$.

Even with the function $A(s)$ unknown, this specific form of
the scattering amplitude is a
non-trivial consequence of the action (45) in the limit of
large $N$.  But to go further and calculate the actual value
of the scattering amplitude we need to restrict ourselves to
low energies.

In the extreme low-energy limit, Eq.~(69) reduces to the
usual low energy Goldstone boson
scattering amplitude,$^8$
\begin{equation}
M(ap,bq\rightarrow
a'p',b'q')\longrightarrow\frac{4}{F_\pi^2}\Big[\delta_{ab}
\delta_{a'b'}\,s+
\delta_{aa'}
\delta_{bb'}\,t+\delta_{ab'}\delta_{a'b}\,u\Big]\;.
\end{equation}
provided we identify the Goldstone boson decay amplitude
$F_\pi$ (equal to $\approx 184$ MeV for pions) as
\begin{equation}
F_\pi=2v\sqrt{N}\;.
\end{equation}
In this low-energy limit, nothing is gained by also taking
$N$ large.

The large $N$ limit does produce some simplification in the
terms of higher order in energy.  According to Eqs.~(67),
(69), and (71), the term in the
Goldstone boson scattering amplitude  of fourth order in
momenta is
\begin{equation}
M^{(4)}(ap,bq\rightarrow a'p',b'q')=-
\frac{N\delta_{ab}\delta_{a'b'}}{2\pi^2 F_\pi^4}s^2\ln(-
s/{\cal M}^2)+{\rm crossed\;terms}
\end{equation}
which may be compared with the exact formula for the terms
in the amplitude of fourth order in
momenta\fnote{\dagger}{This agrees with the result of
reference 5 for the physical case $N=4$.   The term of form
$\delta_{ab}\delta_{a'b'}s^2\ln (-s)$ is proportional to $N-
3$ rather than $N-1$ because it receives contributions from
graphs that do not have index loops as well as from those
that do.}
\begin{eqnarray}
 &&M_{aba'b'}^{(4)} =
\frac{\delta_{ab}\delta_{a'b'}}{F_\pi^4}
\Bigg[ -\;
\frac{N-3}{2\pi^2}~ s^2\ln(-s) -  \frac{1}{12\pi^2}~
(u^2-s^2+3t^2) \ln
(-t) \nonumber \\
& &~~ -~\frac{1}{12\pi^2}~(t^2-s^2+3u^2) \ln(-u)
-cs^2 - c'(t^2+u^2)\Bigg] +
{\rm crossed~terms}\;,\nonumber\\&&{}
\end{eqnarray}
where $c$ and $c'$ are unknown constants.  We see that the
effect of taking the large $N$ limit here is just to
eliminate a few of the terms in Eq.~(73).

Inspection of Eqs.~(67) and (69) shows that not only the
terms in the scattering amplitude of second and fourth order
in a generic momentum $k$, but all the `leading log' terms
of order $k^{2(n+1)}(\ln k^2)^n$ for $n\geq 0$, are uniquely
determined by the first, model-independent term in Eq.~(67),
with no dependence on the coefficients $d_n$ or the
model-dependent functional $H[\lambda]$.  These are just the
model-independent consequences of unitarity and the broken
$O(N)$ symmetry alone, specialized to the case of large $N$.
It is far easier to calculate the leading logarithms by
using
a large $N$ model and then passing to the low energy limit
where the results become model-independent,  as we have done
here, than by evaluating the model-independent leading log
terms for general $N$ and then passing to the limit of large
$N$, but it is still true that to each order in energy there
are only a finite number of diagrams, whether or not we
invoke the large $N$ limit.

\vspace{9pt}
\begin{center}
{\em 2. Unbroken Symmetry Phase}
\end{center}
In this phase the vacuum expectation value of $\phi_r(x)$
vanishes, while $\lambda(x)$ has a vacuum expectation value
$\lambda_0$, given by the solution of the equation
\begin{equation}
\left.\frac{\delta\Gamma[\lambda]}{\delta
\lambda(x)}\right|_{\lambda(x)=\lambda_0}=-
\left.\frac{\partial
V(\lambda)}{\partial\lambda}\right|_{\lambda=\lambda_0}=0\;.
\end{equation}
Here the $O(N)$ symmetry is unbroken, and in the large $N$
limit we have a
degenerate multiplet of scalars with squared mass
$\lambda_0$, so the system will be in this phase only if the
stationary point $\lambda_0$ of $V(\lambda)$ is {\em
positive}.  The $S$-matrix elements for these degenerate
scalars are given in the limit of large $N$ by using the
effective action (60) in the tree approximation; for
instance, the Feynman scattering amplitude is
\begin{equation}
M(rp,sq\rightarrow
r'p',s'q')=\Bigg[\delta_{rs}\delta_{r's'}\Delta(s)
+\delta_{rr'}\delta_{ss'}\Delta(t)+\delta_{rs'}\delta_{r's}
\Delta(u)\Bigg]\;,
\end{equation}
where $\Delta$ is the $\lambda$ propagator, of order $1/N$.

Since $\lambda_0$ is generically of the same order as
whatever characteristic squared mass scale appears in the
functional $\Gamma[\lambda]$, there is no way to use the
model to make useful quantitative  predictions about masses
and
scattering amplitudes in this phase without making special
assumptions about the constants appearing in $H[\lambda]$.
From the large $N$ limit we can only infer conclusions like
Eq.~(75) about the general form of scattering amplitudes.

\vspace{9pt}
\begin{center}
{\em 3. Phase Transition}
\end{center}

As we have seen, in general without knowing all the
constants $c_n$ in the potential $V(\lambda)$ we can say
nothing about the masses in the unbroken symmetry phase
except that they are $O(N)$-degenerate, and without knowing
all the constants $d_n$ in $A(k^2)$ we can say nothing
about the masses in the broken symmetry phase except for the
existence of a massless multiplet of Goldstone bosons.
We can do better in the case where the constants are tuned
so that the system is near the transition between the two
phases.

In the unbroken symmetry phase the system is near the phase
transition if  $\lambda_0$ although non-zero is small.  For
small $\lambda$ Eq.~(62) becomes
\begin{equation}
V(\lambda)\rightarrow
\frac{N\lambda^2\,\ln(\lambda/M^2)}{64\pi^2}+c_1\lambda\;.
\end{equation}
(Recall that $M$ has been chosen to make $c_2$ vanish.)
Condition (74) then becomes
\begin{equation}
c_1=-\frac{\lambda_0\ln(e^{1/2}\lambda_0/M^2)}{32\pi^2}\;.
\end{equation}
This has small positive solutions for $\lambda_0$ as long as
$c_1$ is small and {\em positive}.  On the other hand, in
the broken symmetry phase the system is near the phase
transition if $v_r$ is small, which according to Eq.~(65)
requires that $c_1$ be small and {\em negative}.  Thus there
is a second-order phase transition between these two phases
when $c_1=0$, regardless of the values of the other $c_n$.

Near this phase transition in the broken symmetry phase the
large $N$ approximation allows us to sum  amplitudes to all
orders in the ratio of momenta to the small vacuum
expectation value $v$, provided the momenta are small
compared with all the other mass scales characteristic of
$H[\lambda]$.  Eq.~(68) shows that in this case the
Goldstone boson scattering amplitude has poles at $s=-m^2$,
$t=-m^2$, and $u=-m^2$,
with $m$ given in terms of $v$ by
\begin{equation}
v^2=-m^2A(-m^2)\rightarrow -\frac{m^2\ln(-m^2/{\cal
M}^2)}{32\pi^2}\;,
\end{equation}
indicating the presence of  an unstable light
$O(N-1)$-singlet particle with complex mass $m$.
This is not unexpected; continuity suggests that the $N-1$
Goldstone bosons should be joined near the phase transition
by an additional scalar whose mass must vanish at the phase
transition, in order to allow a smooth transition to the
unbroken symmetry phase, where the $N$ degenerate scalars
become massless at the phase transition.  The mass $m$ given
by Eq.~(78) is complex because this scalar can decay into
Goldstone bosons.

There are
other possibilities: near the phase
transition the unbroken phase could have  a degenerate
multiplet of light
scalars belonging to any representation of $O(N)$ that
contains the $(N-1)$-vector representation of $O(N-1)$, not
necessarily the defining representation.  The
Goldstone bosons of the broken symmetry phase would then be
joined at the phase transition with those massless scalars
that are needed to fill out this representation when $O(N)$
is restored.     Our transformation of the theory has
emphasized the possibility that near the phase transition
the light degenerate multiplet of the unbroken phase forms
an $N$-vector, but of course we do not know that the $c_1$
parameter encountered in this transformed theory {\em is}
small.  To explore other possible types of phase transition,
we would have to transform the theory in other ways, and
then assume that the parameter corresponding to $c_1$ in
those transformed theories is small.

The smallness of $c_1$ opens up a much more powerful role
for the large $N$ approximation.   The expectation value
$<\lambda>$ is then small or zero, and the propagator of
$\phi$ goes as $k^{-2}$ for a four-momentum $k$ which though
small is larger than $<\lambda>$.  On the other hand, the
term in the action of second order in $\lambda$ has a
momentum-independent term which is not small near the phase
transition, so the $\lambda$ propagator must be regarded as
of zeroth order in momenta.  We can count powers of momentum
and/or $\sqrt{c_1}$ in any diagram by dimensional analysis,
with the fields $\phi_r$ and $\lambda$ taken as having
dimensions one and two (in powers of momentum),
respectively.

With this understanding, the action (58) contains one
superrenormalizable (`relevant') term $c_{1{\rm B}}\int
d^4x\,\lambda$; three renormalizable (`marginal') terms
$$
-c_{2{\rm B}}\int d^4x\; \lambda^2\,,~~~-
\mbox{\small$\frac{1}{2}$}\int d^4x\;
\partial_\mu\phi_r\,\partial^\mu\phi_r\,,~~~-
\mbox{\small$\frac{1}{2}$}\int d^4x\;
\lambda\,\phi_r\phi_r\,;
$$
and an infinite number of nonrenormalizable (`irrelevant')
terms, including  terms of second order in derivatives of
$\lambda$ that act as corrections to the
momentum-independent zeroth-order $\lambda$ propagator.
(The subscript B on $c_{1{\rm B}}$ and $c_{2{\rm B}}$
indicates that these are bare couplings, chosen to give
finite  values to the $c_1$ and $c_2$ in Eq.~(62).  To
leading order in $1/N$ there is no renormalization of the
coefficients of $\int d^4x\;
\partial_\mu\phi_r\,\partial^\mu\phi_r$ and $\int d^4x\;
\lambda\,\phi_r\phi_r$; these coefficients are fixed  to be
$-1/2$ by a choice of normalization of $\lambda$ and
$\phi_r$.)   The presence of a superrenormalizable term
prevents an expansion in powers of momenta alone, but near
the phase transition with $c_1$ small, we can expand any
scattering amplitude in powers of the over-all scale of
momenta and $\sqrt{c_1}$.  The leading term in this
expansion is given by Feynman diagrams involving only the
renormalizable and superrenormalizable interactions listed
above.

The presence of the renormalizable interaction $\int
d^4x\;\lambda\,\phi_r\phi_r$ means that there is an
infinite number of multi-loop graphs of leading order in
momenta and/or $\sqrt{c_1}$.  Here the large $N$ limit
offers the huge simplification, of reducing the complete
quantum effective action to the simple form (60), only now
with $H[\lambda]$ containing only terms linear and quadratic
in $\lambda$:
\begin{eqnarray}
\Gamma[\phi,\lambda]&=&-\int
d^4x\;\left[\mbox{\small$\frac{1}{2}$}\partial_\mu\phi_r
\partial^\mu\phi_r+\mbox{\small$\frac{1}{2}$}\lambda
\phi_r\phi_r\right]-Nc_{1\rm{B}}\int d^4x\;\lambda-
Nc_{2\rm{B}}\int d^4x\; \lambda^2\nonumber\\&&+~
\mbox{\small$\frac{1}{2}$}iN\,{\rm Tr}\,\ln (\Box-
\lambda)\;.
\end{eqnarray}

Using Eq.~(79) in the tree approximation gives the terms in
scattering amplitudes of leading order in $1/N$ and in small
momenta and $\sqrt{c_1}$.  For instance, in the broken
symmetry phase the function $A(k^2)$ appearing in Eqs.
(66)--(69) is here
simply given by the first term in Eq.~(67)
\begin{equation}
A(k^2)=\frac{\ln (k^2/{\cal
M}^2)}{32\pi^2}\;.
\end{equation}
For $v\rightarrow 0$ there is just one solution of the
equation $v^2=k^2\,A(k^2)$ with $k^2\rightarrow 0$ (the only
case where Eq.~(80) can be trusted).  This solution has
${\rm Re}\,k^2<0$, and corresponds to an unstable scalar
particle that can decay into pairs of Goldstone bosons.

\vspace{12pt}
\begin{center}
IV. THE EXTENDED $CP^{N-1}$ MODEL: INTEGRATING IN A GAUGE
FIELD
\end{center}

Besides helping us to count factors of $1/N$ and enforcing
constraints on the fields, auxiliary fields are sometimes
introduced in order to enforce a condition of gauge
invariance.  The leading example of this sort is the $CP^{N-
1}$ model,$^4$  which in its original form has an action
given by Eqs.~(11) and (12).  Here we shall consider a class
of extensions of the $CP^{N-1}$ model
in which non-trivial finite results may be obtained in
the limit of large $N$ in four spacetime dimensions.  As we
shall see, just as in its original  two
dimensional version, this model has the remarkable property
that a
long range Coulomb force arises even though no elementary
gauge field is  introduced into the action.

The extended $CP^{N-1}$  models to be considered here
contain a set of $N$ complex scalar fields $u_r$, subject to
the constraint that
\begin{equation}
u_r^\dagger(x)\,u_r(x)=N\;.
\end{equation}
The action is invariant under `gauge' transformations
\begin{equation}
\delta u_r(x)=i\,\epsilon (x)\,u_r(x)\;,
\end{equation}
with $\epsilon(x)$ an arbitrary real infinitesimal function.
For a
soluble model which yields  non-trivial finite results, it
turns out
to be sufficient to take the action in the form
\begin{equation}
I[u]=Nf^2\int d^4x \left(-b+ a_\mu a^\mu\right)+Nf^2F[a,b]
\end{equation}
where
\begin{equation}
a_\mu \equiv  -(i/2N)\Big(u_r^\dagger \partial_\mu u_r-
(\partial_\mu u_r^\dagger) u_r\Big)\;,
\end{equation}
\begin{equation}
b\equiv (1/N)\,(\partial_\mu u_r^\dagger)\,\partial^\mu
u_r\;,
\end{equation}
 and $F[a,b]$ is an arbitrary
$N$-independent Lorentz-invariant perturbatively local
functional, invariant under the transformation induced by
the gauge transformation (82)
\begin{equation}
\delta
b(x)=2a^\mu(x)\partial_\mu\epsilon(x)\;,~~~~~~~~~\delta
a_\mu(x)=\partial_\mu\epsilon(x)\;.
\end{equation}
The first term in Eq.~(83) is a rewritten version of the
action (11), (12) of the original $CP^{N-1}$ model; this
term could have been included in $NF[a,b]$, but it is
convenient to display the kinematic part of the action
explicitly.  In principle we should include all
$SU(N)$-invariant bilinear currents in addition to $a_\mu$
and $b$,
but the effects of other currents are suppressed at small
momenta, and the addition of an arbitrary functional of
$a_\mu$ and $b$ is enough to allow the cancellation of
infinities.   Note that $a^\mu(x)$ is now given by Eq.~(84),
and is {\em not} taken as an  independent field, because we
will need to include terms in  $F[a,b]$ involving
$\partial_\mu a_\nu-\partial_\nu a_\mu$, and we are trying
to see how a Maxwell field can arise without its being put
in from the beginning.  The photon will appear here in quite
a different way.

For the sake of variety, we will take a different approach
to
counting powers of $1/N$ here, which gives the same result
as the functional Fourier transform used in the previous
sections.  We introduce
a {\it pair} of new auxiliary fields $\alpha_\mu(x)$,
$\rho_\mu(x)$ and $\beta(x)$, $\sigma(x)$ for each of the
bilinears $a_\mu(x)$ and $b(x)$ appearing in the action,
writing Eq.~(83) in the equivalent form
\begin{equation}
I=Nf^2\int d^4x\,\Big(-
b+\alpha_\mu\alpha^\mu+\sigma\,(\beta-
b)+\rho^\mu(\alpha_\mu-a_\mu)\Big)+
Nf^2F[\alpha, \beta] \;.
\end{equation}
Integrating out the $\sigma(x)$ and $\rho_\mu(x)$ yields
delta functions which set $\beta(x)=b(x)$ and
$\alpha_\mu(x)=a_\mu(x)$, taking us back to Eq.~(83).
Instead, we shall first integrate out the $\alpha_\mu(x)$
and $\beta(x)$.  Since the terms in (87) that depend on
$\alpha_\mu(x)$ or $\beta(x)$ are simply proportional to $N$
and do not depend on the $u_r$, in the limit of large $N$ we
can set these fields equal to the values at which (87) is
stationary with respect to $\alpha_\mu(x)$ and $\beta(x)$,
giving
\begin{equation}
I=-Nf^2\int d^4x\;\Big((1+\sigma)b+\rho^\mu a_\mu\Big)
+Nf^2G[\rho,\sigma]\;,
\end{equation}
where $G$ is the Legendre transform of $F+\int
d^4x\,\alpha_\mu\alpha^\mu$
\begin{equation}
G[\rho,\sigma]=\Big[F[\alpha,\beta]+\int
d^4x\,\alpha_\mu\alpha^\mu +\int
d^4x\,(\rho^\mu\alpha_\mu+\sigma\beta)\Big]_{\rm staty}
\end{equation}
with the subscript `staty' meaning that we set
$\alpha_\mu(x)$ and $\beta(x)$ equal to values satisfying
conditions that make the quantity in square brackets
stationary:
\begin{equation}
\frac{\delta F}{\delta\alpha^\mu(x)}=-2\alpha_\mu(x)-
\rho_\mu(x)\;,~~~~~~~~
\frac{\delta F}{\delta\beta(x)}=-\sigma(x)\;.
\end{equation}

The Legendre transform of a general perturbatively local
functional is just another general perturbatively local
functional, so we can regard $G[\rho,\sigma]$ as arbitrary,
except for a gauge invariance condition.  Using Eq.~(90) and
the invariance of $F[a,b]$ under the transformation (86), we
easily see that
\begin{equation}
0=\int
d^4x\;\partial_\mu\epsilon\,\Big(2\alpha^\mu\frac{\delta
F[\alpha,\beta]}{\delta\beta}+\frac{\delta
F[\alpha,\beta]}{\delta\alpha_\mu}\Big)
=\int d^4x\;\partial_\mu\epsilon\,\Big(-
2(1+\sigma)\frac{\delta G[\rho,\sigma]}{\delta\rho_\mu}-
\rho^\mu\Big)\;.
\end{equation}
It follows that we can define a new functional
\begin{equation}
G'[\rho,\sigma]\equiv G[\rho,\sigma]+\frac{1}{4}\int
d^4x\;\frac{\rho_\mu\rho^\mu}{1+\sigma}
\end{equation}
that is invariant under the transformations
\begin{equation}
\delta \rho_\mu=-2(1+\sigma)\partial_\mu\epsilon~~~~~~~~~~~
\delta\sigma=0\;.
\end{equation}
This suggests that we should define a gauge field
\begin{equation}
A_\mu\equiv -\frac{\rho_\mu}{2(1+\sigma)}\;,
\end{equation}
which according to Eq.~(93) has the gauge transformation
property
\begin{equation}
\delta A_\mu=\partial_\mu\epsilon\;.
\end{equation}
In the original version of the $CP^{N-1}$ model, $F=0$,
and then Eq.~(90) shows that $\rho_\mu=-2\alpha_\mu=-2a_\mu$
and $\sigma=0$, so Eq.~(94) gives $A_\mu=a_\mu$.  But in the
general case with $F\neq 0$, it is incorrect to identify
$A_\mu$ with $a_\mu$.

We are not yet ready to add a Lagrange multiplier term $-
\int d^4x\,\lambda\,(u^\dagger_ru_r-N)$ and integrate out
the $u_r$ fields, because then we would again encounter an
infinite term proportional to $\int d^4x\,\lambda^2$, and it
is not yet clear how this could be cancelled.  Instead we
will first re-define the fields to introduce an order
parameter, as we did in the previous section for the
non-linear $\sigma$-model.  Define
\begin{equation}
z_r\equiv f\sqrt{1+\sigma}\,u_r\;,
\end{equation}
subject to the constraint
\begin{equation}
z_r^\dagger z_r=Nf^2(1+\sigma)\;.
\end{equation}
The bilinears (84) and (85) then take the form
\begin{eqnarray}
a_\mu &=&\frac{-
i}{2f^2N(1+\sigma)}\Big(z_r^\dagger\partial_\mu z_r-
(\partial_\mu z_r)^\dagger z_r\Big)\\
b&=&\frac{1}{f^2N(1+\sigma)}\partial_\mu z_r^\dagger
\partial^\mu z_r-
\frac{1}{4(1+\sigma)^2}\partial_\mu\sigma\partial^\mu
\sigma\;.
\end{eqnarray}
The action given by Eq.~(88) now may be written
\begin{equation}
I=-\int d^4x\;(D_\mu z_r)^\dagger\,D^\mu z_r
+Nf^2G''[A,\sigma]\;,
\end{equation}
where $D_\mu$ is the gauge-covariant derivative
\begin{equation}
D_\mu z_r\equiv \partial_\mu z_r-iA_\mu z_r\;,
\end{equation}
and $G''$ is another arbitrary gauge-invariant
perturbatively local functional
\begin{equation}
G''[A,\sigma]\equiv \frac{1}{4}\int
d^4x\,\frac{\partial_\mu\sigma\partial^\mu\sigma}{1+\sigma}+
G'[\rho,\sigma]\;.
\end{equation}

Now we may enforce the constraint (97) by introducing a
Lagrange multiplier term
%67
\begin{equation}
-\int d^4x\,\lambda\,\left(z_r^\dagger z_r-
Nf^2(1+\sigma)\right)\;,
\end{equation}
which preserves gauge invariance if we define $\lambda(x)$
to be gauge-invariant.
After integrating out the $\sigma$ field, the action becomes
%68
\begin{equation}
I=-\int d^4x\;(D_\mu z_r)^\dagger\,D^\mu z_r-\int
d^4x\,\lambda z_r^\dagger z_r+NH[A,\lambda]
\;,
\end{equation}
where $H[A,\lambda]$ is yet another Legendre transform
%69
\begin{equation}
H[A,\lambda]=f^2\left[G''[A,\sigma]+\int
d^4x\,(1+\sigma)\lambda\right]_{\sigma=\sigma^\lambda}
\end{equation}
with $\sigma^\lambda(x)$ equal to the  $\sigma(x)$ at which
the quantity in square brackets on the right-hand side of
Eq.~(105) is stationary
\begin{equation}
\left.\frac{\delta G''[A,\sigma]}{\delta
\sigma(x)}\right|_{\sigma=\sigma^\lambda}=-\lambda(x)\;.
\end{equation}
The $z_r$ field is now unconstrained.

We want to calculate the effective action
$\Gamma[z,A,\lambda]$, which in the tree approximation gives
the same result as the sum of all loop and tree graphs
calculated using the action (104).  Following the same
reasoning as in Section 2 (with $z_r$ replacing $\psi_r$,
and $\lambda$ and $A$ replacing $\sigma$),
we find that this is given to leading order in $1/N$ by
\begin{equation}
\Gamma[z,A,\lambda]=-\int d^4x\;(D_\mu z_r)^\dagger\,D^\mu
z_r-\int d^4x\,\lambda z_r^\dagger z_r+\Gamma[A,\lambda]\;,
\end{equation}
where
\begin{equation}
\Gamma[A,\lambda]=iN\,{\rm Tr}\,\ln[D_\mu D^\mu-
\lambda]+N\,H[A,\lambda]\;.
\end{equation}
Gauge invariance and dimensional analysis tell us that the
infinite part of the first term in Eq.~(108) is a linear
combination of
the gauge-invariant functionals $\int d^4x\, \lambda$, $\int
d^4x \,\lambda^2$,
and $\int d^4x \,(\partial_\mu A_\nu-\partial_\nu A_\mu)^2$.
But $H[A,\lambda]$ is an arbitrary perturbatively local
functional, constrained only by invariance under the gauge
transformation (95), so there is no problem in adjusting it
to cancel these infinities.

Like the non-linear sigma-model, the $CP^{N-1}$ model can
exist in several phases, characterized here by different
spacetime-independent vacuum expectation value of the scalar
fields $z_r$ and $\lambda$, with $A^\mu(x)=0$.
In analyzing these phases, we shall make use of the fact
that for $A^\mu(x)=0$ and $\lambda(x)$ constant,
$\Gamma[A,\lambda]$ may be expressed as in Eq.~(63) in terms
of an effective potential $V(\lambda)$
\begin{equation}
\Gamma[0,\lambda]=-{\cal V}_4V(\lambda)\;,
\end{equation}
 with ${\cal V}_4$ the spacetime volume.  The effective
potential here is given by a formula like Eq.~(62)
\begin{equation}
V(\lambda)=\frac{N\lambda^2\,\ln(\lambda/M^2)}{32\pi^2}+N
\sum
_{n=1}^\infty c_n\lambda^n
\end{equation}
with $c_n$ a set of new constant coefficients depending on
$H[0,\lambda]$, and $M$ a new constant that can be chosen to
make $c_2=0$.
(The coefficient in the first term is $1/32\pi^2$ instead of
$1/64\pi^2$ because $z_r$ is complex.)
In analyzing the vector particle mass and the `charge' of
the $z_r$ particles, we will also need to study the term in
$\Gamma[A,\lambda]$ of second order in the
photon field for constant $\lambda(x)$, which gauge
invariance requires  must take the form
 \begin{equation}
\Gamma^{(2)}[A,\lambda]=\frac{N}{2}\int
d^4x\, A^\mu(x)\left(\eta_{\mu\nu}\Box-
\partial_\mu\partial_\nu\right)\,f(-
\Box,\lambda)\,A^\nu(x)\;,
\end{equation}
Evaluating the trace in Eq.~(108) gives
\begin{equation}
f(q^2,\lambda)=-\frac{1}{16\pi^2}\int_0^1 dx\,(1-2x)^2
\ln\left(\frac{\lambda+q^2x(1-
x)}{W^2}\right)+\sum_{n=0}^\infty f_n(q^2)\lambda^n\;,
\end{equation}
where $f_n(q^2)$ are $N$-independent functions of $q^2$
analytic at $q^2=0$, arising from the unknown functional
$H[A,\lambda]$, and
$W$ is another mass parameter, which can be chosen to make
$f_0(0)=0$.

\vspace{9pt}
\begin{center}
{\em  1. Broken Symmetry Phase}
\end{center}

In this phase $z_r(x)$ has a non-vanishing vacuum
expectation value $\sqrt{N}v_r$ while the vacuum expectation
values of $\lambda(x)$
and $A^\mu(x)$ both vanish,
which requires that
\begin{equation}
\left.\frac{\partial V(\lambda)}{\partial
\lambda}\right|_{\lambda=0}=-Nv^2
\end{equation}
where $v^2\equiv v^*_rv_r$.  Since $\Gamma\propto N$, $v_r$
is $N$-independent.
For small constant $\lambda$, the effective potential (110)
is
\begin{equation}
V(\lambda)\rightarrow\frac{N\lambda^2\,\ln(\lambda/M^2)}
{32\pi^2}+ N\,c_1\lambda
\end{equation}
Hence condition (113) gives
\begin{equation}
c_1=-v^2\;.
\end{equation}
In analyzing the degrees of freedom in this phase, it is
very convenient to eliminate the scalar--vector mixing in
Eq.~(107) by adopting unitarity gauge, in which
Im$(v_r^*z_r)=0$.  Taking $v_N=v$ real and $v_i=0$ for
$i=1,\cdots N-1$, this means that $z_N$ is real, while the
$z_i$ are still complex.  The $z_i$ are massless Goldstone
boson fields, while $z_N$ has the same sort of mixing with
$\lambda$ that we saw in the previous section; the terms in
the action of second-order in $\lambda$ and/or $z_N-
\sqrt{N}v$
have a coefficient matrix given by
\begin{eqnarray}
&&{\cal D}_{NN}(k^2)=k^2\;,~~~~ {\cal
D}_{\lambda\lambda}(k^2)
=NA(k^2)\;, \nonumber\\&&
{\cal D}_{N\lambda}(k)={\cal D}_{\lambda N}(k)=\sqrt{N}v
\end{eqnarray}
where now
\begin{equation}
A(k^2)=\frac{\ln (k^2/{\cal
M}^2)}{16\pi^2}+\sum_{n=1}^\infty f_n(k^2)^n\;.
\end{equation}
with $f_n$ yet another set of model-dependent constants;
and ${\cal M}$ is a constant chosen so that the term $f_0$
in the sum is absent.
The scalar mass $m$ is given by the condition that this have
a zero determinant at $k^2=-m^2$:
\begin{equation}
-m^2 A(-m^2)=v^2\;.
\end{equation}
Without further information
 about the functional $H[0,\lambda]$, we
cannot tell whether there actually is a massive scalar in
the
spectrum of $z_N$ and $\lambda$.

To study the vector
particles in this phase, we note that according to
Eq.~(107), the  term in
$\Gamma[v,A,0]$ of second order in the photon field  is
given in this phase by
\begin{equation}
\Gamma^{(2)}[\sqrt{N}v,A,0]=-Nv^2\int
d^4x\,A_\mu(x)A^\mu(x)+\Gamma^{(2)}[A,0]
\end{equation}
where $\Gamma^{(2)}[A,0]$ is defined by Eq.~(111).    There
is
a vector particle of mass $\mu\neq 0$ if
\begin{equation}
\mu^2f(-\mu^2,0)=2v^2\;.
\end{equation}
Without special assumptions about $H[A,0]$, it is not
possible to tell this has a solution, much less to calculate
the vector boson mass $\mu$.     But it is clear that any
massive
scalar or vector particles would have to be unstable,
because they could decay into the Goldstone bosons $z_i$.

\vspace{9pt}
\begin{center}
{\em  2. Unbroken Symmetry Phase}
\end{center}
In this phase $\lambda$ has  a  non-zero vacuum expectation
value $\lambda_0$, satisfying the condition
\begin{equation}
\left.\frac{\partial V(\lambda)}{\partial
\lambda}\right|_{\lambda=\lambda_0}=0\;,
\end{equation}
which allows $z_r(x)$ as well as $A^\mu(x)$ to have
vanishing expectation values.  Eq.~(107) shows that in this
phase the $z_r$ have squared mass $\lambda_0$, so
$\lambda_0$ must be positive.
 The photon propagator in momentum space equals
$$\Delta_{\mu\nu}(k)=\frac{\eta_{\mu\nu}}{k^2\,
N\,f(k^2,\lambda_0)} + {\rm gauge~~terms}\;,$$ where `gauge
terms' denotes gauge-dependent
terms proportional to $k_\mu k_\nu$.  Because the  $z_r$ for
$\lambda_0\neq 0$ have a finite mass the function
$f(k^2,\lambda_0)$ is analytic at $k^2=0$, so  the photon
here is massless.  Also the renormalized gauge field is
$\sqrt{Nf(0,\lambda_0)}A^\mu$, so the  $z_r$ charge is
$1/\sqrt{Nf(0,\lambda_0)}$, and is hence of order
$1/\sqrt{N}$.  Without making special assumptions about the
functional $H[A,\lambda]$ it is impossible to say anything
more about the values of the $z_r$ squared mass $\lambda_0$
or the $z_r$ charge.

\vspace{9pt}
\begin{center}
{\em  3. Phase Transition}
\end{center}

As in the case of the non-linear $\sigma$-model, we can
obtain more detailed results when the model is near a phase
transition between the broken and unbroken symmetry phases.
In the unbroken symmetry phase, the model is near this phase
transition if the $\lambda_0$ satisfying Eq.~(121) is
positive and small.  In this case  Eqs.~(110) and (121) give
\begin{equation}
c_1=-\frac{\lambda_0\ln(e^{1/2}\lambda_0/M^2)}{16\pi^2}\;
\end{equation}
which has positive small solutions for $\lambda_0$ as long
as $c_1$ is small and positive.  On the other hand, in the
broken symmetry phase the model is near the phase transition
if $v$ is small, which according to Eq.~(115) requires that
$c_1$ is small and negative.  Thus there is a second-order
phase transition at $c_1=0$, irrespective  of the values of
other parameters.

To analyze the low-momentum limit near a phase transition,
we note that the action (104) contains a single
superrenormalizable term $-Nc_{1{\rm B}}\int d^4x\,\lambda$;
four strictly renormalizable terms
\begin{eqnarray*}
&&-Nc_{2{\rm B}}\int d^4x\,\lambda^2\;;~~~-
\mbox{\small$\frac{1}{4}$}NZ\int d^4x\; (\partial_\mu A_\nu-
\partial_\nu A_\mu)^2 \nonumber\\&&
-\int d^4x\;(D_\mu z_r)^\dagger\,D^\mu z_r\;;~~~-\int
d^4x\,\lambda\, z_r^\dagger z_r \;;
\end{eqnarray*}
and an infinite number of non-renormalizable terms.  (The
subscript B again indicates bare values, adjusted to give
finite values to $c_1$ and $c_2$ in Eq.~(110).)  In the
limit where $c_1$ and all momenta are small, we can ignore
the non-renormalizable interactions, and calculate
scattering amplitudes by using the quantum effective
interaction (107) in the tree approximation, now with
\begin{eqnarray}
\Gamma[A,\lambda]&=&iN\,{\rm Tr}\,\ln[D_\mu D^\mu-
\lambda]-Nc_{1{\rm B}}\int d^4x\,\lambda-Nc_{2{\rm B}}\int
d^4x\,\lambda^2 \nonumber\\&&-
\mbox{\small$\frac{1}{4}$}NZ\int d^4x\; (\partial_\mu A_\nu-
\partial_\nu A_\mu)^2 \;.
\end{eqnarray}
This tells us that for example that the potential
$V(\lambda)$ is given by Eq.~(114);  that for constant
$\lambda(x)$ the function
$f(q^2,\lambda)$ appearing in the formula (111) for the term
in $\Gamma[A,\lambda]$ of second order in the vector field
is here
\begin{equation}
f(q^2,\lambda)=-\frac{1}{16\pi^2}\int_0^1 dx\,(1-2x)^2
\ln\left(\frac{\lambda+q^2x(1-
x)}{W^2}\right)\;;
\end{equation}
and that the function $A(k^2)$ appearing in the
scalar two-point function (116) is
\begin{equation}
A(k^2)=\frac{\ln (k^2/{\cal
M}^2)}{32\pi^2}\;.
\end{equation}

As an example of the use of these results, let's look more
closely at the properties of
the particles  near the phase transition.  In the unbroken
symmetry phase for small $\lambda_0$, Eq.~(124) gives
\begin{equation}
f(0,\lambda_0)\rightarrow -
\frac{N}{48\pi^2}\ln\left(\frac{\lambda_0}{W^2}\right)
\end{equation}
This is positive but diverges for $\lambda_0\rightarrow 0$,
so that the $z_r$ charge $1/\sqrt{N\,f(0,\lambda_0)}$
vanishes
at the phase transition.  In the broken symmetry phase
the vector boson mass is determined by the function
$f(q^2,0)$, which for $q^2\rightarrow 0$ is given by
Eq.~(124) as
\begin{equation}
f(q^2,0)\rightarrow -
\frac{1}{48\pi^2}\left[\ln\left(\frac{q^2}{W^2}\right)-
\frac{8}{3}\right]\;.
\end{equation}
Eq.~(120) for the vector
boson squared mass $\mu^2$ has a single solution that
vanishes as $v\rightarrow 0$, indicating the presence of a
single light vector particle.     Also,   in the broken
symmetry phase near the phase transition, Eq.~(118) for the
scalar mass $m$ takes the form
\begin{equation}
\frac{-m^2\,\ln (-m^2/{\cal M}^2)}{32\pi^2}=v^2\;.
\end{equation}
This has one solution for $m^2$ that vanishes as
$v\rightarrow 0$, indicating the presence of single massive
but light scalar particle.  These solutions for  $\mu^2$ and
$m^2$ both have positive real part but are complex,
reflecting the fact that both of these particles are
unstable, because they can decay into pairs of Goldstone
bosons.

\vspace{12pt}
\begin{center}
V. DYNAMICAL GAUGE BOSONS: A REMARK
\end{center}

The $CP^{N-1}$ model has attracted much attention because
of the appearance of a massless gauge boson in a theory
involving only scalar fields.  It is important to recognize
that this phenomenon does not depend on the existence of the
gauge symmetry (82), or indeed on any of the symmetry
properties of the action.

This can be seen by a very general argument.$^9$  Consider a
theory that is invariant under a gauge group $G$, with
various matter multiplets forming various representations of
$G$.  Suppose that one of these multiplets consists of
scalar fields, some of which have vacuum expectation values
that
completely break the gauge symmetry.  Integrate out the
massive gauge vector bosons in unitarity gauge.  We then
have a perturbatively local effective field theory, with no
hint of the original gauge invariance.  It seems pretty
clear that if we allow arbitrary interactions in the
original theory, then in this way we obtain a completely
general effective field theory of the remaining fields.  But
this procedure can be reversed, so {\em out of any effective
field theory with no gauge symmetry and possibly no global
symmetry we can obtain a theory with any broken gauge
symmetry.}

The point is that a spontaneously broken gauge symmetry in
itself has no predictive power.$^{10}$  Of course, it can
have
plenty of predictive power if the gauge coupling is {\em
weak}, but for this we have to fine-tune the parameters in
the action.  In the $CP^{N-1}$ model studied in the previous
section, this fine-tuning is achieved by the condition that
$c_1$ is small.

To illustrate the possibility of constructing a broken gauge
symmetry in an effective field theory that has no symmetry
to begin with, consider a theory of Dirac fields $\psi_i(x)$
with an action of form
\begin{equation}
I[\psi]=-\int
d^4x\,\sum_i\overline{\psi_i}\gamma^\mu\partial_\mu\psi_i
-G[\psi]
\end{equation}
where  $G[\psi]$
is an essentially arbitrary perturbatively local functional
of the fermion fields.  We can choose $G[\psi]$ so that this
action has {\em no} internal symmetries --- if we like, not
even fermion conservation.  This action can be obtained by
integrating out a vector field $A_\mu(x)$ in the action
\begin{eqnarray}
I[\psi,A]&=&-\int
d^4x\,\sum_i\overline{\psi_i}\gamma^\mu\partial_\mu\psi_i
-G'[\psi]- \frac{M^2}{2}\int d^4x\,A_\mu A^\mu
\nonumber\\&&+\int d^4x\,A_\mu j^\mu-\frac{1}{4}\int
d^4x\,F_{\mu\nu}F^{\mu\nu}
\end{eqnarray}
where $M$ is an arbitrary mass parameter;
$F_{\mu\nu}(x)\equiv \partial_\mu A_\nu-\partial_\nu A_\mu
$; and
\begin{eqnarray}
G'[\psi]&\equiv & G[\psi]-\frac{1}{2}\int d^4x\,
j_\mu(x)\frac{1}{M^2-\Box}j^\mu(x)\nonumber\\&&-
\frac{1}{2M^2}\int
d^4x\,\partial_\nu j^\nu(x)\frac{1}{M^2-\Box}\partial_\mu
j^\mu(x)\;.
\end{eqnarray}
where $j^\mu(x)$ is the current
\begin{equation}
j^\mu\equiv\sum_i q_i\overline{\psi_i}\gamma^\mu\psi_i\;,
\end{equation}
with $q_i$ an arbitrary set of real parameters.
As long as $M\neq 0$, $G'$ is still perturbatively local.
The action (130) can be obtained from another action
\begin{eqnarray}
I[\psi,A,u]&=&-\frac{M^2}{2}\int d^4x\, \left|\partial_\mu
u-iA_\mu u\right|^2-\int
d^4x\,\sum_i\overline{\psi_i}\gamma^\mu\Big(\partial_\mu-
iq_i A_\mu\Big)\psi_i \nonumber\\&&-\frac{1}{4}\int
d^4x\,F_{\mu\nu}F^{\mu\nu}+G'[\psi']
\end{eqnarray}
with $u$ a scalar field constrained by $|u|^2=1$,
and
\begin{equation}
\psi_i'\equiv \psi_i\,u^{-q_i}\;.
\end{equation}
The action (133) is invariant under the gauge transformation
\begin{equation}
\psi_i(x)\rightarrow
e^{iq_i\alpha(x)}\psi_i(x)\;,~~u(x)\rightarrow
e^{i\alpha(x)}u(x)\;,~~A_\mu(x)\rightarrow
A_\mu(x)+\partial_\mu\alpha(x)\;,
\end{equation}
so the action (130) can be obtained from (133) by adopting
the unitarity gauge, in which $u=1$.  (The action (133) is
also perturbatively local, because  in deriving perturbation
theory, we expand around $u=1$ rather than $u=0$.)  Yet
there is no trace of this gauge invariance or even global
invariance in the action (129) with which we started.

\pagebreak

\begin{center}
{\bf ACKNOWLEDGEMENTS}
\end{center}

I am grateful for helpful discussions with S. Coleman, J.
Distler, and V. Kaplunovsky.
This research was
supported in part by the
Robert A. Welch
 Foundation and NSF Grant PHY 9511632.

\begin{center}
{\bf REFERENCES}
\end{center}
\begin{enumerate}
\item For a review of the large $N$ limit in
various contexts, see S. Coleman,
in {\em Aspects of Symmetry} (Cambridge University Press,
Cambridge, 1985): Chapter 8.

\item The earliest reference known to me for the $O(N)$
linear and non-linear $\sigma$-models is M. Gell-Mann and M.
L\'{e}vy, {\it Nuovo Cimento}
{\bf 16}, 705 (1960).  The $O(N)$ linear $\sigma$-model was
studied in the large $N$ limit in statistical mechanics by
H. E. Stanley, {\it Phys. Rev.} {\bf 176}, 718 (1968); K.
Wilson, {\it Phys. Rev.} {\bf D7}, 2911 (1973); and in
four-dimensional relativistic quantum field theories by
L. Dolan and R. Jackiw, {\it Phys. Rev.} {\bf D9}. 3320
(1974); H. J. Schnitzer, {\it Phys. Rev.} {\bf D10}, 2042
(1974); S. Coleman, R. Jackiw, and H. D. Politzer, {\it
Phys. Rev.} {\bf D10}, 2491 (1974); and many later authors.
I don't know who first studied the non-linear $\sigma$
model in the large-$N$ limit.

\item D. Gross and A. Neveu, {\em Phys. Rev.} {\bf D10},
3235 (1974).

\item H. Eichenherr, {\em Nucl. Phys.} {\bf B146}, 215
(1978); V. Golo and A. Perelomov, {\em Phys. Lett.} {\bf
79B}, 112 (1978); A. D'Adda, M. L\"{u}scher, and P. Di
Vecchia, {\em Nucl. Phys.} {\bf B146}, 63 (1978); {\bf
B152,}, 125 (1979); E. Witten, {\em Nucl. Phys.} {\bf B149},
285 (1979); H. Haber, I. Hinchcliffe, and E. Rabinovici,
{\em Nucl. Phys.} {\bf B172}, 458 (1980); M. Bando, T. Kugo,
and K. Yamawaki, {\it Phys. Rept.} {\bf 164}, 217 (1988).

\item S. Weinberg, {\it Physica} {\bf 96A}, 327
(1979).

\item See, e. g., R. de M. Koch and J. P. Rodrigues,
Witwatersrand preprint hep-th/9605079 (1996).

\item S. Coleman and E. Weinberg, {\it Phys. Rev.} {\bf D7},
1888 (1973).

\item S. Weinberg, {\it Phys. Rev. Lett.} {\bf 17},
616 (1966).

\item V. Kaplunovsky, private communication.

\item See, e. g., S. Weinberg, {\em The Quantum Theory of
Fields II: Modern Applications} (Cambridge University Press,
Cambridge, 1996): p. 318.
\end{enumerate}
\end{document}